\documentclass[fleqn,10pt]{wlscirep}
\usepackage[utf8]{inputenc}
\usepackage[T1]{fontenc}
\title{Self-Sustained Non-Equilibrium Co-existence of Fluid and Solid States in a Strongly Coupled Complex Plasma System}

\author[1,*]{M.G Hariprasad}
\author[1]{P. Bandyopadhyay}
\author[2,3,+]{V. S. Nikolaev}
\author[2,3]{D. A. Kolotinskii}
\author[4]{S. Arumugam}
\author[1]{G. Arora}
\author[1]{S. Singh}
\author[1]{A. Sen}
\author[3,2]{A. V. Timofeev}

\affil[1]{Institute for Plasma Research, A CI of Homi Bhabha National Institute, Bhat, Gandhinagar, Gujarat 382428, India}
\affil[2]{Moscow Institute of Physics and Technology, Dolgoprudnyi, Moscow region, 141701, Russia}
\affil[3]{Joint Institute for High Temperatures, Russian Academy of Sciences, Moscow, 125412, Russia}
\affil[4]{Department of physics,  Sikkim Manipal Institute of Technology, Majitar, Sikkim- 737136}
\affil[*]{hari.prasad@ipr.res.in}
\affil[+]{vladislav.nikolaev@phystech.edu}

\begin{abstract}
A complex (dusty) plasma system is well known as a paradigmatic model for studying the kinetics of solid-liquid phase transitions in inactive condensed matter. At the same time, under certain conditions a complex plasma system can also display characteristics of an active medium with the micron-sized particles converting energy of the ambient environment into motility and thereby becoming active. We present a detailed analysis of the experimental complex plasmas system that shows evidence of a non-equilibrium stationary coexistence between a cold crystalline and a hot fluid state in the structure due to the conversion of plasma energy into the motion energy of microparticles in the central region of the system. The plasma mediated non-reciprocal interaction between the dust particles is the underlying mechanism for the enormous heating of the central subsystem, and it acts as a micro-scale energy source that keeps the central subsystem in the molten state. Accurate multiscale simulations of the system based on combined molecular dynamics and particle-in-cell approaches show that strong structural nonuniformity of the system under the action of electostatic trap makes development of instabilities a local process. We present both experimental tests conducted with a complex plasmas system in a DC glow discharge plasma and a detailed theoretical analysis.
\end{abstract}

\begin{document}

\flushbottom
\maketitle
\thispagestyle{empty}

Interest in  model systems that permit a study of phase transitions in condensed matter started in the 1930s with the theory of the Wigner crystal \cite{wigner}. Model systems provide significant insights for  furthering  our understanding of mechanisms of formation and growth of new phases and scenarios of phase transitions. In the scope of this problem, macroscopic systems with a high degree of non-ideality, such as colloidal suspensions in aqueous solutions \cite{activecolloids} and dusty plasmas \cite{dustyplasmas}, are of a special interest. Their properties can be studied in detail on a particle-resolved level.

In this paper, the focus is on dusty plasmas which are actively employed as a paradigm for studies of phase transitions \cite{phasedusty, melzer2012phase}, collective wave excitations \cite{wavesdusty}, self-organization \cite{fortovmorfill, polyakov2015synergetics}and transport processes \cite{transportdusty}.  Micron-sized particles (dust) injected into a gas discharge plasma acquire a high negative electric charge of the order of $10^4$~$e$ by collecting the highly mobile electrons in the plasma. These highly charged dust particles can constitute a strongly coupled (non-ideal) component of the system and can form organized structures like crystals. These structures can be studied at the particle (kinetic) level by video-microscopic methods. 
Although dusty plasmas are often considered as toy models for the study of inactive condensed matter,  under certain conditions it mimics active matter \cite{activematterreview} and reveals similarity with such objects as cells in tissues \cite{cellsintissues}, suspensions of bacteria \cite{bacteriasuspensions} and even the flockings of birds \cite{flocksbirds}. The common principle uniting all of these seemingly different systems is the specific mechanism of energy exchange between the system and the environment.  Energy is injected into such systems at the level of each individual particle and is converted into the motion of particles. This leads to an intrinsic out-of-equilibrium behavior of the system and its departure from the action-reaction symmetry \cite{ivlev2015statistical} at the microscopic level. The source of injected energy might be, for example, chemical interaction, as in self-phoretic Janus colloids \cite{januscolloids}. 
 
In dusty plasmas, there are several mechanisms that might make dust particles behave similar to agents in active matter. These mechanisms include the effects of ion and neutral shadowing \cite{khrapak2009basic} and the "rocket effect" \cite{nosenko2010laser}. There are also experimental studies of Janus particles in a gas discharge plasma which become active due to the action of photophoretic force from the illuminating laser \cite{janusdusty, petrovactive}. Another mechanism leading to energy conversion under certain conditions is the plasma-specific wake effect\cite{wakeeffect, wakeeffect1, wakeeffect2}. The wake effect arises due to the action of electric field that compensates for gravity acting on dust microparticles. This electric field also exerts a strong vertical ion flow which is perturbed by a highly charged particle of dust. The radial distribution of the electrostatic potential in the wake region around each dust particle is no longer spherically symmetric \cite{matthews2020dust, piel2017molecular, hutchinson2007computation, miloch2010charging, melzer2014connecting, ivlev2017instabilities}. Asymmetry of the electrostatic potential and momentum exchange with the plasma flow lead to the nonreciprocal character of particle interactions in the dust subsystem. Particles absorb energy and momentum from the flowing plasma and then transport it from the local scale to the larger scales. 

\par
There are several examples of specific mechanisms in complex plasmas that might lead to an ordered-disordered structure coexistence in the dust subsystem although such coexistence has not been studied in detail. {Nosenko \textit{et al.} \cite{torsion} performed an experiment where spontaneous formation of spinning pairs of particles (torsions), is studied in a mono-layer complex plasma crystal by reducing the discharge power at constant neutral gas pressure. Number of torsions are found to be increasing with decrease in the discharge power. At lower gas pressures, the torsions are preceded by mode coupling instability (MCI). A crystal-fluid coexistence is produced in the system by MCI, but the transient nature of the coexistence is not discussed. Moreover, structural or thermodynamical parameter measurements are also not carried out to explore the coexistence.} In another experiment, Uchida \textit{et al.} \cite{vortex_flows} produced a two-dimensional dust vortex flows in an ordered complex plasma structure. The  vortex structures are produced by the effect of an asymmetry of ion drag force. The ordered structure is found to be disturbed by the dust flow, and  disordered structure is produced in the vortex region which leads to an ordered-disordered structure coexistence in the system. Moreover, in a very recent molecular simulations study by Qiu \textit{et al.} \cite{shock_coexistence}, a two-dimensional Yukawa solid and liquid separation after the shock propagation is reported. After the propagation of compressional shocks, the structure and dynamics of the post-shock region are investigated. When the compressional speed is significantly higher than a first threshold value, the post-shock region melts completely. However, when this compressional speed is lower than a second threshold value, which is smaller than the first threshold, the post-shock region is found to be in the solid state. Whereas, at a compressional speed value between first threshold and second threshold, the post-shock region clearly exhibits the coexistence of the solid close to the compressional boundary and the liquid in the other part. Further, the averaged kinetic temperature in the post-shock region shows a spatial variation and is attributed to the dynamical heterogeneity of the 2D Yukawa systems. \par

In the present work we focus on the possible coexistence regime which is due to the plasma-specific mechanism of non-equilibrium melting in complex plasmas. This mechanism is caused by the wake effect and may arise mainly due to two types of instabilities.  In multilayer systems, with at least two layers, the so-called ``Schweigert" instability \cite{melzer1996structure, schweigert1996alignment, schweigert1998plasma} develops in the form of driven horizontal oscillations of particles. It is due to the non-reciprocal interaction of dust particles located at different levitation heights. In a single-layered (mono-layer) quasi-2D system of dust particles  another type of instability can arise due to the coupling of the horizontal motion of dust particles to their vertical motion. In this case of a mode coupling instability \cite{couedel2009first, couedel2011wave} particles oscillate with approximately equal amplitudes and frequencies in both the horizontal and vertical directions \cite{melzer2014connecting}. The conditions for the onset of MCI are determined by the strength of the vertical confinement, the density of the system and the neutral gas damping rate.
 
In most experimental works considering non-equilibrium melting of dust structures under the action of instabilities only melting of the entire structure is reported. The intermediate state where coexisting phases can be observed has not been studied experimentally. However, as soon as the threshold for the instability onset in dusty plasmas is density dependent in most cases, phase coexistence can be expected if the collection of particles is spatially nonuniform \cite{prxnonuniformity}. Spatial nonuniformity of structural and dynamical characteristics reveals in conditions of many experiments with dusty plasmas due to the action of central electrostatic confinement which keeps the system stable \cite{hari_transition, dpexcrystal, nikolaevtimofeev}. Such central trap can only be compensated by an inter-particle interaction force gradient and subsequent density gradient. The recent theoretical study of MCI-induced melting in the finite-size dust monolayer demonstrates that, if density gradient in the system is strong enough, then the peripheral part of the structure might stay ordered and coexist with the molten central region where MCI is active \cite{nikolaev2021nonhomogeneity}. Moreover, Schweigert instability may also trigger a coexisting phase if the instability is driven locally in the system. This scenario demonstrates the applicability of phase coexistence concept to the collections of "active Brownian particles" in dusty plasmas where the particles in the locally molten region acquire energy from plasma through a non-equilibrium process. In the present paper, we present such a system where a non-equilibrium solid-fluid coexistence exists and the dust particles in fluid phase are driven active by the ion wake mediated non-reciprocal interaction.

We provide the detailed experimental and theoretical study of coexistence of solid and fluid states in a strongly coupled system of dust particles in a gas discharge plasma. In the experiment, the system of dust particles has complex geometry: multilayer 3D structure in the central region and single-layered quasi-2D structure at the periphery. Detailed profiles of kinetic temperature are given with particle-resolved precision. It is shown that a strong temperature gradient is present in the system: "hot" central 3D-liquid part coexists with "cold" peripheral 2D-crystalline part. The co-existing parts are in the state of dynamical equilibrium, thus the co-existence is stationary and self-sustained. The heat flux from the central part to the peripheral region is supported by the continuous energy input from the wake-mediated particle interaction. Theoretical description of the system is based on the multiscale approach to simulations: while the dynamics of dust particles is described using the Newton equation, the pairwise inter-particle interaction is calculated by the particle-in-cell method accounting ion-neutral collisions. In simulations the onset of instability in the central region along with the absence of instability at the periphery is demonstrated.
\section*{\label{sec:exp}Phase coexistence experiment in complex plasmas}
 \label{sec:exp}

\subsection*{\label{sec:experiment} Experimental set-up}

\begin{figure*}
\centering
\includegraphics[scale=0.5]{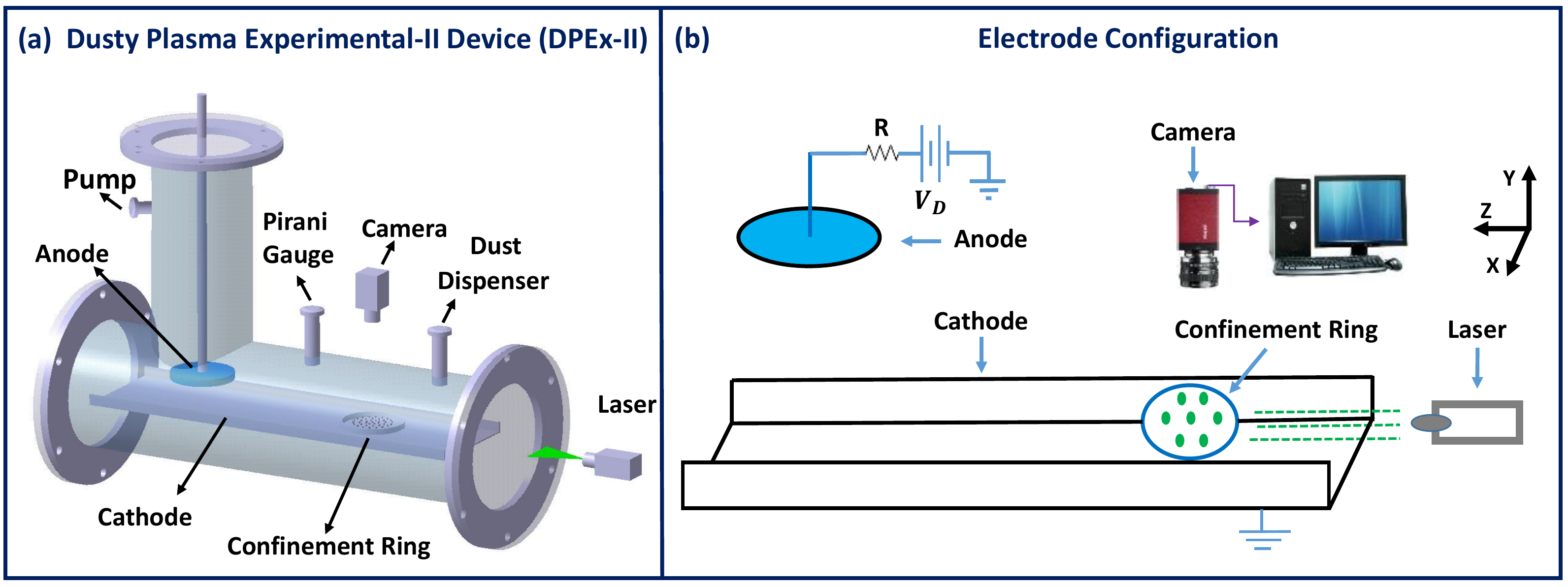}% 
\caption{\label{fig:fig1} Schematic diagram of the (a) DPEx device and (b) electrode configuration with diagnostic tools.}
\end{figure*}

All the experiments presented in this work have been performed in the Dusty Plasma Experimental-II (DPEx-II) device \cite{dpexii}, which has  an $L$-shaped vacuum vessel made of pyrex-glass consisting of two chambers as shown in Fig.~\ref{fig:fig1}(a). The plasma and the complex (dusty) plasma are produced in the primary chamber, and the auxiliary chamber provides connections to different subsystems for evacuating the chamber and producing and characterizing the plasma and the complex plasma. A base pressure of 0.1~Pa is initially achieved in the experimental chamber by a rotary pump, and  Argon gas is then flushed several times through it to remove the impurities. The working gas pressure is finally set at $\sim$ 8 Pa using a mass flow controller. In the DPEx-II device, an asymmetrical electrode configuration consisting of a circular-shaped stainless steel anode, and a long tray-shaped cathode are employed. Fig.~\ref{fig:fig1}(b) represents the schematic diagram of this asymmetrical electrode configuration. A DC glow discharge Argon plasma is produced between the electrodes by applying a DC voltage of 450~V. The plasma current is estimated by measuring the voltage drop across a current limiting resistor of 2~k$\Omega$ that is connected in series with the power supply. The plasma parameters such as the electron temperature, plasma density, plasma potential, etc., are measured using single and double Langmuir probes as well as emissive probes. For this discharge condition, the plasma density and the electron temperature are found to vary over a range of 0.8--2$\times 10^{15}$~m$^{-3}$ and 2--4~eV, respectively. Mono-dispersive melamine-formaldehyde (MF) particles of diameter 10.66 $\mu$m are then introduced in the plasma by a dust-dispenser to form the complex plasma. These dust particles in the plasma environment get negatively charged by accumulating more electrons (due to their higher mobility) than ions and levitate in the cathode sheath region due to a balance between the downward gravitational force and the upward electrostatic force resulting from the cathode sheath. The dust particles get confined horizontally due to the sheath-electric field around a confinement ring placed on the cathode and form a circular-shaped mono-layer crystal . The micron-sized particles are illuminated by a green laser and a CCD camera is used to capture the Mie-scattered light. In our experiments, the width of the laser used for illuminating the particles is comparatively smaller than the inter-layer separation, which allows us to scan the particles layer by layer. However, the particles which move randomly due to their stochastic thermal motion in the liquid state are also considered in the analysis. A sequence of images is stored in a computer for further analysis. IDL and MATLAB-based software are employed for tracking the individual particles over time to study their dynamics. \par 

A hexagonally ordered complex plasma structure is observed when the neutral gas pressure and the discharge voltage are set at 8 Pa and 450 V, respectively. Production of such a stable mono-layer complex plasma crystal in the cathode sheath region of a DC glow discharge plasma was recently reported by Hariprasad \textit{et al.}\cite{dpexcrystal}. At that given discharge voltage, the plasma density decreases with a reduction of neutral gas pressure, which causes the sheath around the confining ring to thicken. As a result, the confinement area for the dust particles shrinks, which causes a few particles at the center of the mono-layer to move below the mono-layer  as shown in Fig.~\ref{fig:fig2}(b)\cite{hari_transition} . These particles then interact strongly with the particles that reside in the mono-layer and form a fluid-like structure in the central region. Thus, with the decrease of neutral gas pressure, a purely crystalline state changes to an ordered-disordered coexisting state, which remains stable over time. Fig.~\ref{fig:fig2}(a) shows a typical image of the coexistence of ordered-disordered states at a neutral gas pressure of $\sim$ 7.3~Pa. At that specific discharge condition, the central region of the complex plasma exhibits a disordered fluid-like structure surrounded by an ordered crystalline structure. It is also found that the particles in the center move randomly and don't possess an equilibrium position whereas they arrange themselves into an ordered state  at the periphery and oscillate around their equilibrium position. The structural and the thermodynamical properties of these distinct regions of  coexisting states differ significantly, and they will be delineated in detail in the upcoming sections.

\subsection*{Coupling parameter estimation}
\label{subsection:gamma}
The Coulomb coupling parameter ($\Gamma$) is defined as the ratio of the potential energy of the dust particles with the kinetic energy \cite{ikezi1986, vaulina_melting} and it determines the phase state of the complex plasma system. In the past, many studies have been  carried out to investigate the dependency of complex plasma system on the Coulomb coupling parameter \cite{ikezi1986,langevindynamics,dpexcrystal}. Ikezi \textit{et al.} \cite{ikezi1986} predicted that the Coulomb system exhibits an ordered structure when the Coulomb coupling parameter crosses a threshold value of $\Gamma \sim 172$ if the definition of the coupling parameter is made via the Wigner-Seitz radius. For the screened Coulomb system, Vaulina and Khrapak\cite{vaulina_melting} predicted a threshold value of the effective coupling parameter $\sim$ 106 using the definition via the mean inter-particle distance. The difference between the numerical values $172$ and $106$ for an unscreened system is explained by the ratio of the Wigner-Seitz radius to the mean inter-particle distance.  Knapek \textit{et al.}\cite{langevindynamics} introduced a new measurement technique to estimate the Coulomb coupling parameter and the dust temperature from the position information of dust particles.

In this approach, dust particles are assumed to be distinguishable classical particles and obey Maxwell-Boltzmann distribution. In local equilibrium, when the interaction of these dust particles with the plasma as well as with individual neutral atoms is, on average, balanced by neutral friction, the dynamics of individual particles in the lattice can in principle be described by a Langevin equation \cite{langevindynamics}. In such a case, one can write down the probability distribution \cite{langevindynamics} for each lattice cell as, \par
%%%%%%%%%%%%%%%%%%%%
\begin{eqnarray}
\centering
P(r,v)\propto exp\left[-\frac{m{(v-<v>)}^2}{2T}-\frac{m{\Omega_E}^2r^2}{2T}\right],
\end{eqnarray}
%%%%%%%%%%%%%%%%%%%%
 with $T$ being the particle temperature, $\Omega_E$ the Einstein frequency and $m$ the dust particle mass.  The standard deviation of the velocity distribution and of the displacement distribution independently yield the dust temperature and the coupling parameter, respectively. The standard deviation of the velocity distribution is given by ${\sigma_v}= \sqrt{\frac{T}{m}}$ and the standard deviation of the displacement distribution is given by ${\sigma_r}=\sqrt{\frac{T}{m\Omega_E^2}}=\sqrt{\frac{\Delta^2}{\Gamma_{eff}}}$. In the present set of experiments, the Coulomb coupling constant is estimated from the knowledge of inter-particle distance and standard deviation of displacement distribution as $\Gamma_{eff}=\left[\frac{\Delta}{\sigma_r}\right]^2$.

In our experiments, we estimated the coupling parameter and the dust temperature by employing the above mentioned technique  \cite{dpexcrystal}. From the position information of the dust particles, the displacement distribution is obtained along with the standard deviation in both ordered and disordered phases \cite{dpexcrystal}. Inter particle distances are calculated from the particle positions to estimate the coupling parameter \cite{dpexcrystal}. The value of the effective coupling parameter in the ordered peripheral region is estimated to be $\sim$ 210, while in the disordered structure at the center the value turns out to be $\sim$ 35. Both the values and the statement of phase coexistence in the observed system are in agreement with the threshold values obtained by Vaulina and Khrapak\cite{vaulina_melting} and also with our earlier experiments~\cite{dpexcrystal, hari_transition}. The crystal-fluid phase coexistence of the complex plasma system is further confirmed by the Voronoi diagram based structural analysis.

\subsection*{\label{sec:Analysis}Voronoi diagram based  structural analysis}

\begin{figure*}
\centering
\includegraphics[scale=0.5]{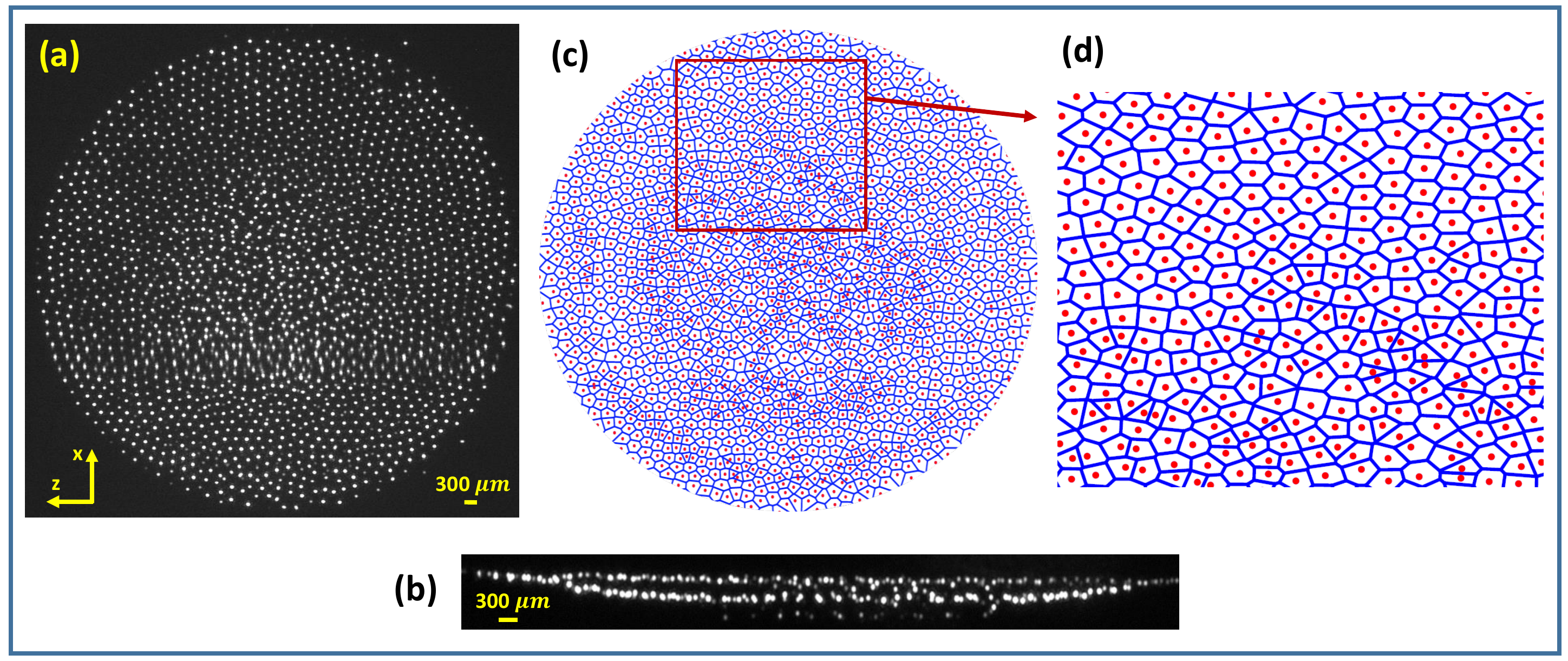}% 
\caption{\label{fig:fig2} (a) Top and (b) side view of the phase coexisting complex plasma system. (c) Corresponding Voronoi Diagram of Fig.~\ref{fig:fig2}(a). (d) Zoomed view of phase coexisting boundary region.}
\end{figure*}

The coexisting structure is further analysed using a Voronoi diagram \cite{voronoi}, which segregates the plane with $n$ points into convex polygons such that every polygon contains exactly one generating point, and each point in a given polygon comes closer to its generating point compared to others. This diagram provides the basic unit cell of the structure under investigation with the information of coordinate number and the defects present in the system. Fig.~\ref{fig:fig2}(c) depicts the Voronoi diagram of the phase coexisting structure of Fig.~\ref{fig:fig2}(a). It is seen from the Voronoi diagram (depicted in Fig.~\ref{fig:fig2}(c)) that the outer portion of the structure consists of hexagonally oriented unit cells and forms a crystalline state. In contrast, the central portion of the Voronoi diagram consists of completely disordered polygons that are filled with defects and therefore essentially depict a fluid state. The amount of disorder and the number of defects become maximum at the center, and they reduce as one moves to the periphery. Fig.~\ref{fig:fig2}(d) presents a zoomed view of the crystal-fluid coexisting boundary region in which the top and the bottom portion of the complex plasma correspond to crystalline and fluid states, respectively. There does not exist a rigid boundary that separates the coexisting phases; instead, the defects reduce towards the periphery. The transformation from the disordered fluid to the crystalline phase occurs within a distance of approximately four to five inter-particle distances. Hexagonal cells dominate in the peripheral portion with some crystal defects, which are inevitable in a finite plasma crystal due to confinement boundary effects. Thus, the Voronoi diagram provides a detailed overview of the structural nature of the crystal-fluid  coexistence state in a complex plasma system.

\subsection*{\label{sec:temp}Non-equilibrium nature of the coexistence state}

An important concern regarding a phase coexistence structure is the nature of its thermodynamic state. Generally, in equilibrium systems, the temperatures of the two coexisting phases remain the same \cite{equilibrium} whereas, in a non-equilibrium phase coexistence state, the phases can exhibit different temperatures \cite{activematter}. To determine the thermodynamic nature of our experimental phase coexistence state we have estimated the temperatures at different locations of the dusty plasma to cover  the regions of the two phases. For this the trajectories of individual particles at various locations are examined from a sequence of recorded images in order to estimate their average velocities and the dust temperature is estimated as explained in the previous section. By assuming the dust particles to obey a Maxwell-Boltzmanian distribution, the average temperature of the dust particles is estimated from the experimentally obtained velocity distribution. The full width at half maximum ($T= m{\sigma_v}^2$, where, $\sigma_v$ is the standard deviation of the velocity distribution and $m$ is the mass of the dust particles) of the distribution function gives the average temperature of the particles \cite{langevindynamics,dpexcrystal}.  
\begin{figure}
\centering
\includegraphics[scale=1.0]{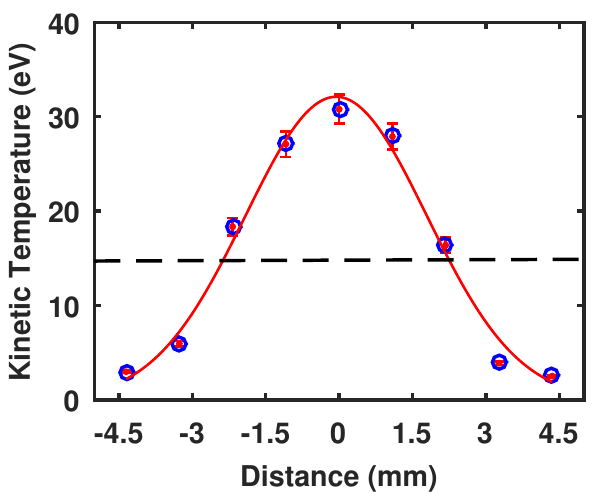}% 
\caption{\label{fig:fig3} Temperature profile of the Phase coexisting structure. Zero of the X-axis is the center of the structure and the dotted line shows the melting temperature.}
\end{figure}

Fig.~\ref{fig:fig3} displays the spatial temperature profile of the complex plasma which has two coexisting phases.  The \lq 0'  location refers to the center of the complex plasma. Fig.~\ref{fig:fig3} shows that the dust temperature has a maximum value ($\sim$ 30~eV) in the fluid phase (at center), which then reduces by almost ten folds ($\sim$ 3~eV) in the crystalline phase (at the periphery). The melting temperature is depicted in the Fig.~\ref{fig:fig3} as a dotted line, to distinguish the coexisting phases more clearly. This drastic difference of temperatures between the two coexisting phases essentially implies the non-equilibrium nature of the system. The temperature profile along the crystal axis looks like the shape of a Gaussian distribution. This is because the dust density is higher at the center where the ion wake driven instability heats up the system more efficiently. The dust density reduces when one goes away from the center and it results in the deduction of the strength of instability and as well the particle temperature. This gives rise to the non-uniformity in the temperature profile. Moreover, the heat produced at the central region diffuses towards the periphery. At the crystal-fluid interface the heat exchange continues and the crystal temperature rises even though there are no particles beneath. Such a smooth transition of dust temperature was also observed in laser heated dusty plasma systems in the past \cite{nosenko2008heat}. Interestingly, in our experiments the coexisting structure is found to be stable over time and both the phases continue to remain at different temperatures even if the system is left for a longer time. Since the complex plasma system is dissipative in nature, there has to be an energy source that heats the central region and keeps the system in a self-sustained non-equilibrium phase coexistence state. One possibility is the existence of an instability mechanism, such as a Schweigert instability \cite{instability1, instability2, Prx},  that can heat the central portion of the complex plasma system  through the ion wake mediated non-reciprocal interaction between the dust particles in the upper and lower layers. This will be briefly  discussed in the later part of this paper. \par
{In the past, Nosenko \textit{et al.} \cite{torsion} performed an experiment by reducing the RF power which results in the change of horizontal and vertical potential trapping of a two-dimensional dust crystal resides in the sheath region of plasma. Due to this change, a few particles leave the crystal plane and get trapped at the bottom and exhibit either circular or even more complex motion in the ion wake regions. It finally leads to the melting of a larger area. They confirmed this as the non-reciprocal Schweigert effect. They described it as once a torsion forms it becomes a source of disturbance in the plasma crystal. Through the particle interactions, later this disturbance or instability redistributes a part its energy that it received from the flowing ions to the surrounding lattice. Moreover, at a particular pressure of 12~Pa, they observed mode coupling instability in the system which they described as  another process of melting. Regarding the different scenarios they concluded that vertical pair formation and MCI are two competing scenarios for the crystal response to weakening vertical confinement and the neutral gas pressure seems to be a key factor in determining which process takes place first. These two cases are easily distinguished by the very different particle trajectories involved. In our experiments, the triggering mechanism for the melting is the Schweigert effect as of Nosenko's first set of observations. Whereas in addition, we performed a detailed analysis of the non-equilibrium coexistence structure produced by partial melting of the crystal. Also we did not observe the MCI induced melting, since the melting and coexistence occurred in a multi layered system in our experiments rather than a single layer system where MCI instability can excite.} \par

%%%%%%%%%%%%%%%%%%%%
%%%%%%%%%%%%%%%%%%%%

\section*{\label{sec:theory}Multiscale simulations of phase coexistence in complex plasmas}

In order to study the underlying mechanism of phase coexistence described in the experimental section, we perform a series of MD simulations for the system of $2500$~dust particles similar to the one observed in the experiment. The description of interaction between dust particles in MD simulations is based on the explicitly calculated distribution of electrostatic potential around dust particles. The prodecure of potential calculation allows to account for the wake effects in the system\cite{wakeeffect, wakeeffect1, wakeeffect2}. Many important plasma parameters required for the potential calculation, such as the temperature and concentration of electrons, the ion flow velocity and the electric field gradient,can be only approximately estimated from conducted experimental measurements. For this reason, the spatial distibution of electrostatic potential is calculated using several sets of parameters which fall into the experimentally estimated range. The procedure allowing to calculate the wake potential self-consistently via the PIC approach \cite{hutchinson2002ion, hutchinson2006collisionless, miloch2012dust, changmai2020particle} is given in Supplementary Materials. The typical view of the spatial distribution of the electrostatic potential around a dust particle under considered conditions is shown in Supplementary Fig. S1. The comparison of potential profiles at different values of chosen parameters is given in Supplementary Fig. S2. The obtained collection of results for different wake potentials allows to identify the processes leading to non-equilibrium separation of phases in the experimental system.

\begin{figure}[ht]
\centering
\includegraphics[width=\linewidth]{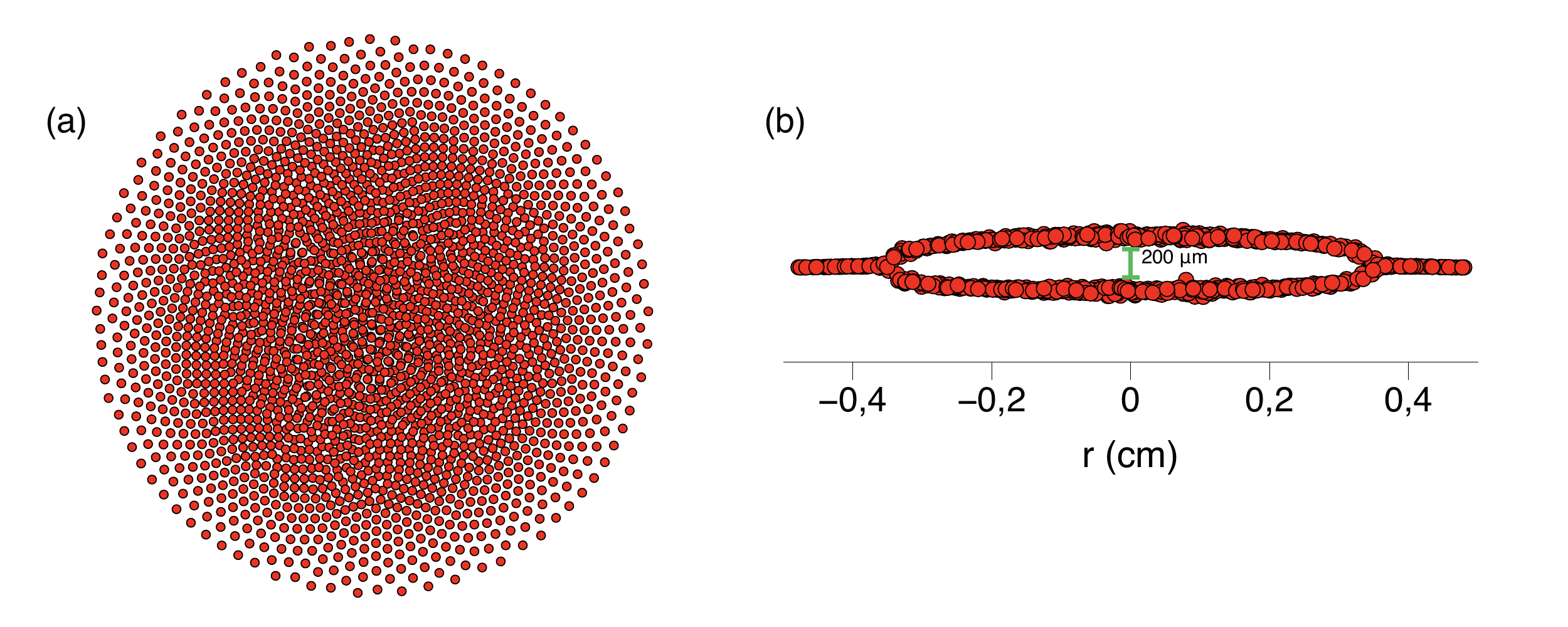}
\caption{Typical (a) top and (b) side view of the dusty plasma structure in MD simulations in this work.}
\label{fig:fig5}
\end{figure} 

\begin{figure}[ht]
\centering
\includegraphics[width=\linewidth]{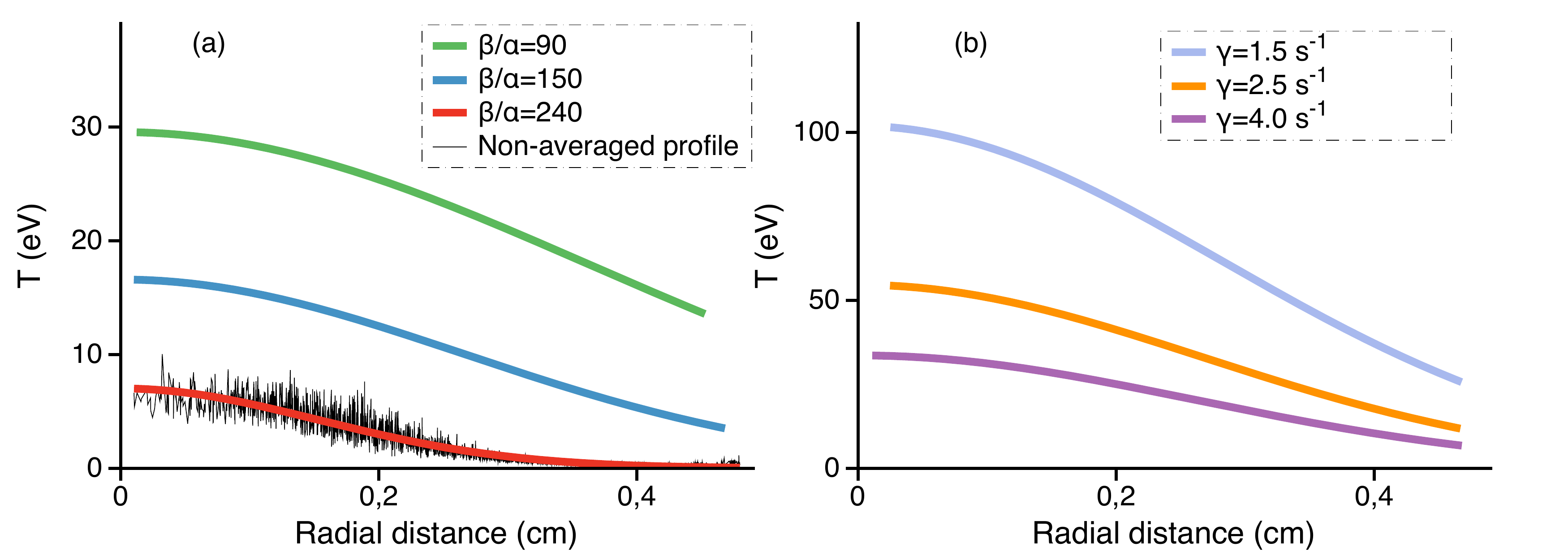}
\caption{The radial profiles of in-plane kinetic temperature in the dusty plasma structure for different (a) values of $\beta$ when $\gamma$ is fixed and equals $7.6$~s$^{-1}$; (b) values of $\gamma$ when $\beta/\alpha$ is fixed and equals $150$. Parameters of the MD and self-consistent wake potential calculations are $Q=15000~e$, $\alpha=0.035$~cgs units, $v_{\rm fl}=v_{\rm B}$.}
\label{fig:fig6}
\end{figure}

For MD simulations, mass $m$ and diameter $D$ of dust particles in the model system are chosen equal to the experimental values. The value of charge of a dust particle $Q$ corresponds to the value of grain charge in the particular wake potential calculation. The time step of molecular dynamics equals $10^{-3}/\omega_{\rm pd}$, where $\omega_{\rm pd} = \sqrt{Q^2/md^3}$ is the plasma-dust frequency and $d$ is the distance between dust particles.

\begin{figure}[ht]
\centering
\includegraphics[width=.4\linewidth]{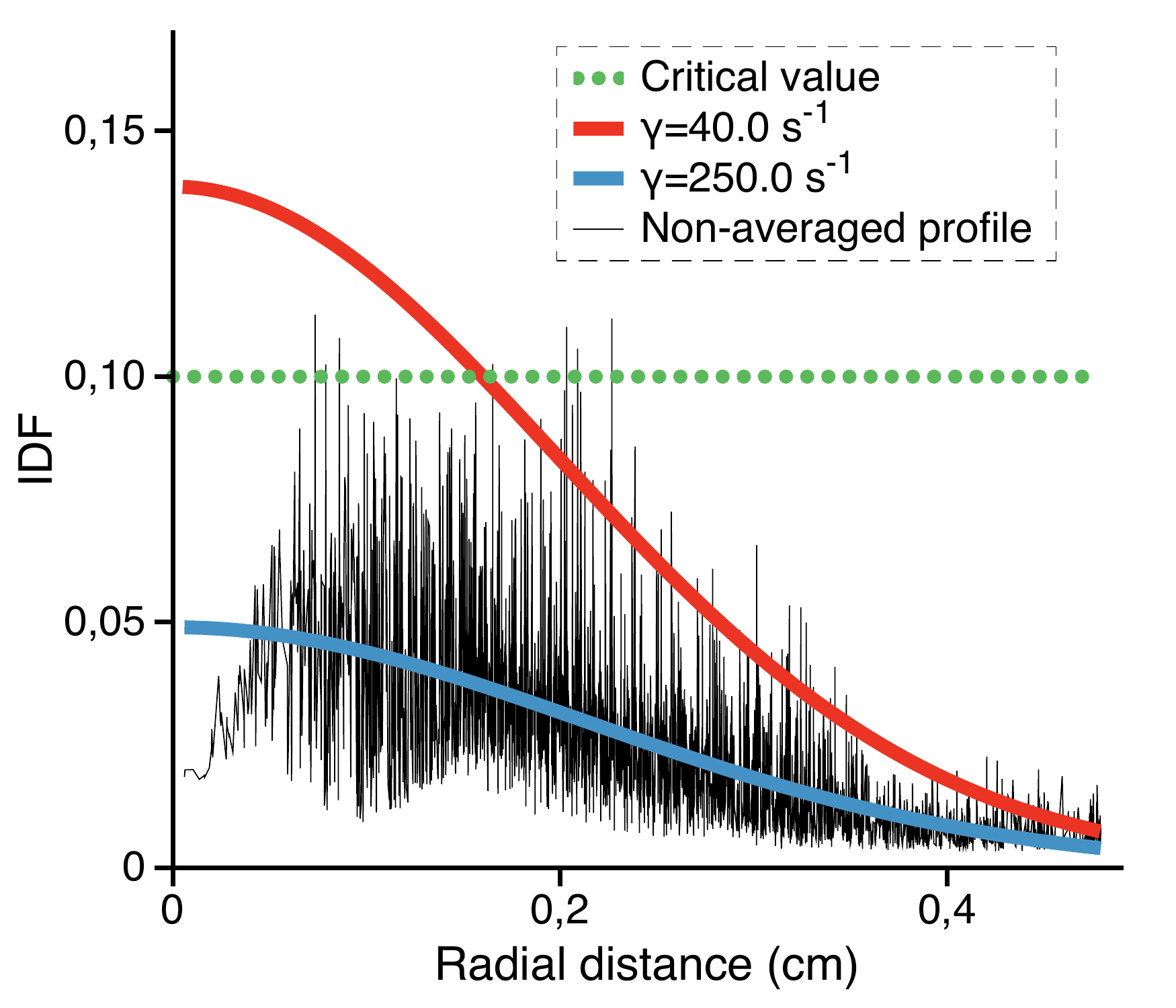}
\caption{IDF radial profile in the dusty plasma system with the parameters $Q=15000~e$, $v_{\rm fl}=v_{\rm B}$, $\alpha=0.035$~cgs units and $\beta/\alpha=230$ at different values of $\gamma$. If the value of IDF is higher than the critical value, the subsystem is molten.}
\label{fig:fig7}
\end{figure}

We include the additional central potential into simulations that confines the likely charged particles and keeps the system stable:
\begin{equation}
	U_{\rm trap} = \frac{1}{2} [\alpha(x_i^2 + y_i^2) + \beta z_i^2],
\end{equation}
where ${\mathbf{r}}_i=(x_i, y_i, z_i)$ is the radius-vector of an $i$-th particle, $\alpha$ and $\beta$ are the horizontal and the vertical trap parameters, respectively. At each value of the particle charge, the values of trap parameters are found to obtain the same system size and geometry as in the experiment. In all simulations, $\alpha$ is of the order of $10^{-2}$~cgs units, $\beta$ is of the order of $1.0$~cgs units. It is important to note that the strength of vertical confinement $\beta$ plays a crucial role in the development of instabilities in a dusty plasma system \cite{melzer2014connecting, ivlev2017instabilities}. For this reason, the value of $\beta$ is varied in each simulation to observe the effect on the system properties.

The overall equation of motion that is solved numerically for the $i$-th dust grain has the form
\begin{eqnarray}
	m \ddot{\mathbf{r}}_i = -Q \nabla U_{\rm{trap}} - 
	- Q \nabla \sum_{j} \varphi(r_{ij}) - m\gamma \dot{\mathbf{r}}_i + \mathbf{L}_i,
\end{eqnarray}
where the last two terms are mathematically equivalent to the action of Langevin thermostat at a room temperature \cite{turq1977brownian}. The term $- m\gamma \dot{\mathbf{r}}_i$ describes the friction of dust particles in the viscous gas environment. Its effect can be significant because the value of the damping factor $\gamma$ can control the development of instabilities in the system. Dependence of the damping factor on the neutral gas pressure can be calculated using the following approximate formula \cite{fortovmorfill}:
\begin{equation}
	\gamma = \frac{2\sqrt{2\pi}}{3}  \frac{P D^2}{v_{\rm n} m},
	\label{eqn:damping}
\end{equation}
where $v_{\rm n}$ is the thermal velocity of neutral gas atoms. In order to study how instabilities develop in the simulated dust structures, the value of $\gamma$ is varied around the value that corresponds to the pressure of neutral gas in the experiment.

Each simulation with unique values of $Q$, $\alpha$, $\beta$ and $\gamma$ starts with randomly positioned particles which interact via the reciprocal Yukawa potential. Under the action of confinement and mutual Yukawa interaction with each other, likely charged particles move to their equilibrium positions. Their kinetic temperature at this moment is equal to the one of the applied Langevin thermostat. Then the interaction of particles is substituted with the chosen wake potential calculated for the charge $Q$ and the ion flow velocity $v_{\rm fl}$. System dynamics comes to the steady state, and then principal data is calculated. We use cutoff radius for the interaction $10$ times higher than the average inter-particle distance in the structure.

The typical view of the simulated system is given in Fig.~\ref{fig:fig5}. It has the same radial size as the experimental structure and a similar value of inter-layer spacing in the central part. Depending on the value of $\beta$, the simulated structure can contain a different number of layers in the central region. At high values of $\beta$, the entire system is one-layered. With the decrease of $\beta$, additional layers form in the center of the structure and the single-layered peripheral subsystem decreases in size. This is an expected behavior for a dust system which transforms from a pancake-like structure into a spherical one at $\beta/\alpha=1$ \cite{yukawaballs, vaulina2009dynamics}. In the range $\beta/\alpha=50\div300$ the central region comprises two or three layers which is the case of interest similar to the experimental system. For this reason, $\beta$ is varied in this range.

Varying $\beta$ and $\gamma$ effects dynamics of the system and allows to observe phase co-existence in several structures with different grain charges $Q$. The important feature of the observed phase co-existence in all of the structures is the presence of strong temperature gradient between the center and the periphery of a structure. Under temperature, we mean the parameter of the velocity distribution which has a Maxwellian profile. The temperature gradient is due to the development of a wake-induced instability in the central region after the formation of the second layer. Intensive energy release has a local character and reaches the periphery of the system in the form of the heat flux. As the heat flux is not enough for the melting of the peripheral part, we observe non-equilibrium phase coexistence of a 3D central "liquid" and a 2D peripheral "crystal" in the simulated system.

The radial profiles of in-plane kinetic temperature of dust particles in the structure are given in Fig.~\ref{fig:fig6} for the values $Q=15000~e$ and $\alpha=0.035$~cgs units. The temperatures are averaged over multiple runs. It can be seen that the increase of $\beta$ leads to the decrease of temperature profile in the given system. The increase of $\beta$ compresses the system in the vertical direction and strengthens the vertical confinement.  The variation of $\gamma$ effects the phase co-existence process in the expected manner: the lower $\gamma$, the more intensive heating of the central region. With the decrease of $\gamma$ the phase boundary moves into the crystalline section of the structure until it melts entirely.

In order to demonstrate phase co-existence quantitatively, we rely on the parameter of inter-particle distance fluctuation (IDF) whose applicability is recently demonstrated for a dust monolayer \cite{nikolaev2021nonhomogeneity}. This parameter can be used to define local phase state in a system where phases co-exist. Formulated for finite-size systems, IDF can be applied both to dust structures \cite{baumgartner2007shell, boning2008melting} and nanoparticles \cite{zhou2002distance}. It has advantages over the popular Lindemann and local order parameters \cite{klumov2009structural} in the considered case because it allows calculation for a system including both 3D and quasi-2D sections. At the same time, its application is simple: the subsystem is supposed to lose order when the numerical value of IDF is above $0.1$. 

As soon as our system contains $2500$~particles, IDF can be used directly without modifications \cite{melzer2012phase}. It is calculated by the formula:
\begin{equation}
	\Delta = \frac{2}{N(N-1)} \sum_{1 \leq i < j}^{N} \sqrt{ \frac{\langle r_{ij}^2 \rangle}{\langle r_{ij} \rangle ^2} -1 },
\end{equation}
where $r_{ij}$ is distance between particles $i$ and $j$; $N$ is the number of particles used to calculate the parameter. Note that IDF is calculated locally, not over the entire structure but over small subsystems containing up to $15$~particles. This allows to track phase state locally both in the central region and at the periphery.

\begin{figure}[ht]
\centering
\includegraphics[width=.4\linewidth]{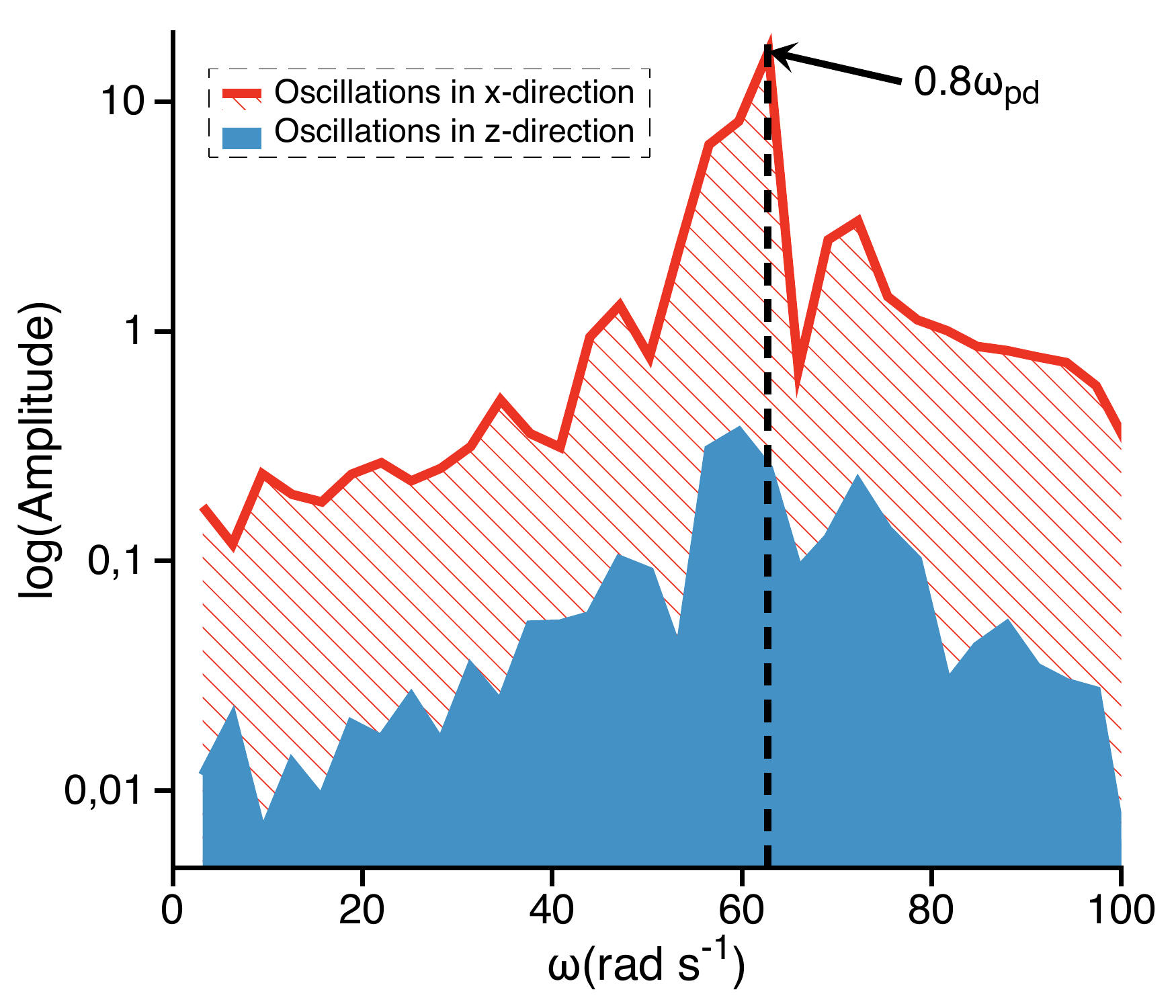}
\caption{Spectrum of particle oscillations in the central region of the dusty plasma system at the values of parameters $Q=15000~e$, $v_{\rm fl}=v_{\rm B}$, $\alpha=0.035$~cgs units, $\beta/\alpha=150$, $\gamma=31.0$~s$^{-1}$. The spectrum is constructed from the velocity autocorrelation function. The vertical black line denotes the value of oscillation frequency $0.8\omega_{\rm pd}$  which allows to identify the Schweigert-type instability. $x$-direction is chosen arbitrarily in the horizontal plane.}
\label{fig:fig8}
\end{figure}

\par
The radial profiles of IDF are shown in Fig.~\ref{fig:fig7} both for the states with and without phase coexistence in the system with $Q = 15000$~e, $\alpha = 0.035$~cgs units and $\beta/\alpha = 230$. The IDF parameter (called ``Berry parameter’’ or even ``Lindemann parameter’’ in a few works) was originally employed for nanoparticles of inert gas atoms. It was shown that when its value calculated for the entire particle reaches $0.1$, it indicates melting of the structure. Later the parameter was adapted for small clusters of Yukawa particles in the modified form via the variance of its block averaged value. Recently it was shown that the original form of IDF can be employed in the local approximation for a monolayer of dust particles. In this case, IDF is calculated for each separate crystalline cell and then averaged over the subsystem of interest. This local form of IDF has a nonuniform radial profile in a monolayer and roughly indicates melting of its shells at the same critical value of $0.1$. In this work, this value is also applicable to the central section of the observed experimental system. The radial distribution function constructed separately for the central section of the structure indicates the loss of long-range order when the local value of IDF is above $0.1$. The main advantage of IDF over the classical Lindemann parameter in the considered case is that it is calculated from relative particle positions and is not susceptible to fluctuations arising in the finite system. The type of instability that drives the observed phase co-existence is discussed in the next section. \par

\subsection*{\label{sec:theory4} Analysis of instabilities observed in simulations}

While geometry of the considered dusty plasma structures in MD simulations is identical, we observe different behavior of the simulated systems with different values of grain charge and wake potentials. The behavior depends on several factors. First, the value of grain charge determines the strength of interaction and the range of melting temperatures in the system. Second, the important factor is the principal nonhomogeneity of the structure: it is dense and two-layered in the central region and rarefied single-layered at the periphery. For this reason, necessary conditions for the instabilities that provide the energy input into the system are satisfied only in the bilayer central subsystem and have a local character. Finally, the type of instability that develops in the central region is determined by the inter-layer distance and the structure of a wake potential.

The stability principles for bilayer complex plasmas are well studied in case of extended systems \cite{melzer2014connecting, ivlev2017instabilities}. There are two types of instabilities that might arise in the bilayer system. The dynamical instability, also known as the Schweigert instability \cite{melzer1996structure, schweigert1996alignment, schweigert1998plasma}, is of the oscillatory type. It arises due to a wave mode that is pumped by the energy of the ion flow and grows exponentially until damping limits its growth. This mode can be suppressed by sufficiently strong neutral gas friction.  It has an important fingerprint that allows to identify its development: in case of the Schweigert instability, particles mainly oscillate in the in-layer direction. Vertical oscillations become negligibly small compared to the horizontal ones. Oscillations in the horizontal direction occur at a frequency which typically lies in the range $0.6\omega_{\rm pd}-1.1 \omega_{\rm pd}$ for the conditions of dusty plasma experiments.

Another type of instability that might arise in the bilayer system is the structural instability which is not of the oscillatory type \cite{ivlev2017instabilities}. It is due to the state of neutral equilibrium when the equilibrium positions of particles no longer correspond to the ground state of the system. The extended motion of particles is present in the structure for any damping. Note that both the dynamical and the structural instability can be mixed.

In our MD simulations, we observe that both types of instabilities can develop locally only in the bilayer central region of the structure. To identify the type of instability, we employ the following technique. Starting from the phase co-existing state, the friction coefficient $\gamma$ is increased to cool down the central region of the system. This step allows to remove low-frequency modes which are observed in the spectrum of the liquid state and to identify the mode responsible for the arise of instability. 

The structural instability is observed in the central region at the values $Q=15000~e$, $\alpha=0.035$~cgs units, $\beta/\alpha=230$. In this case we find that the disordered, flow-like motion of particles is present in the bilayer central region even when the friction coefficient $\gamma$ is many times higher than in the experiment. There is no fingerprint of an oscillatory instability type in the spectrum of particle oscillations.

\begin{figure}[ht]
\centering
\includegraphics[width=.4\linewidth]{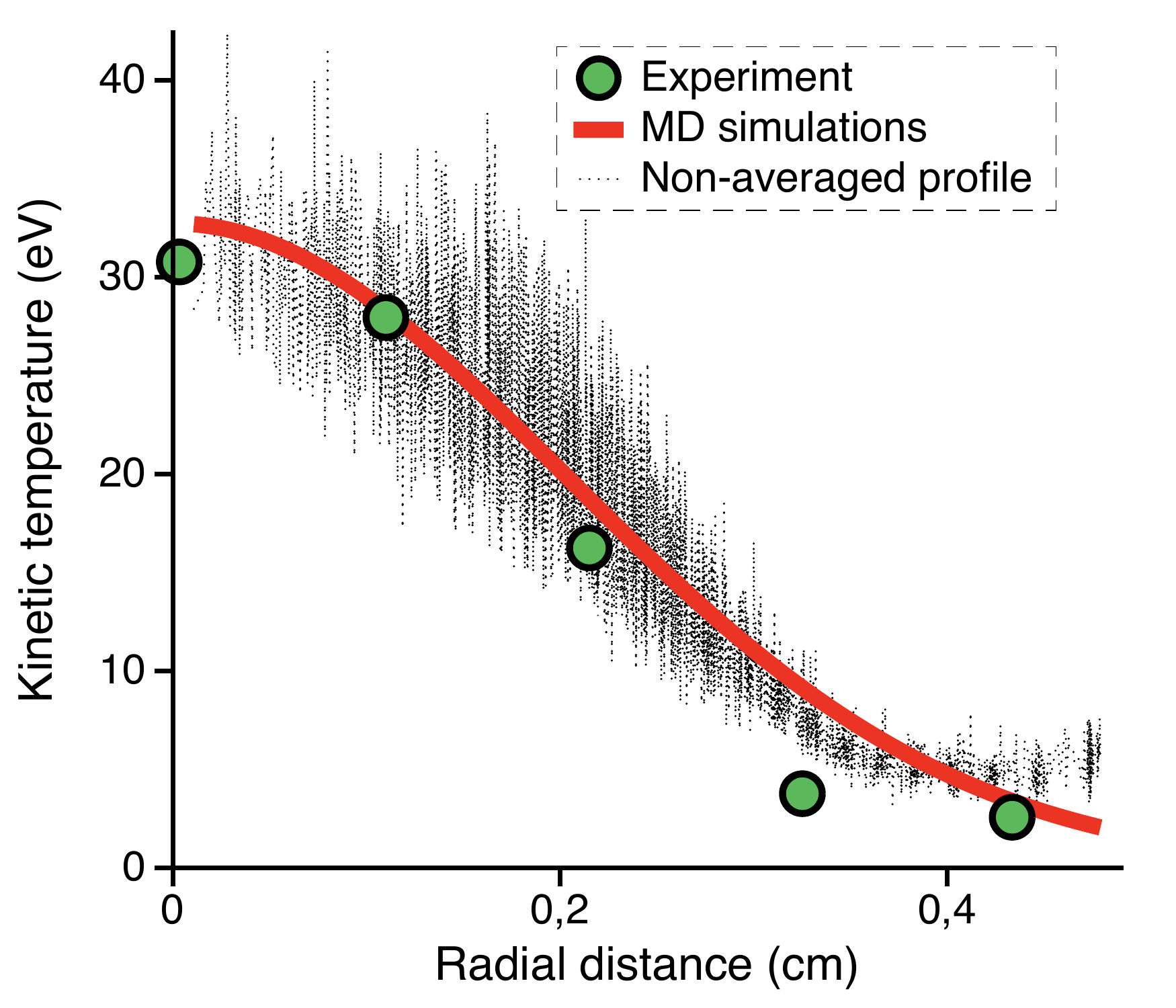}
\caption{Comparison of the radial profiles of kinetic energy obtained in this work experimentally and in MD simulations. The simulated dusty plasma system has the parameters $Q=15000~e$, $v_{\rm fl}=v_{\rm B}$, $\alpha=0.035$~cgs units, $\beta/\alpha=230$, $\gamma=1.8$~s$^{-1}$.}
\label{fig:fig9}
\end{figure}

In most simulations it is the Schweigert-type instability that is responsible for the heating of the central region. An example of the oscillation spectrum for the parameters $Q=15000~e$, $\alpha=0.035$~cgs units, $\beta/\alpha=150$, $\gamma=31.0$~s$^{-1}$ is given in Fig.~(\ref{fig:fig8}). The main peak of in-plane oscillations and the ratio of horizontal to vertical amplitudes identifies the dynamical instability of Schweigert type. It develops locally in the multi-layered central region and fades behind the interface between the 3D and the 2D sections of the structure. The single-layered peripheral subsystem might only become unstable due to the mode coupling instability which does not develop in the considered conditions. For this reason, the following physical picture of the phase co-existence process can be formulated.

When particles in the central region organize into the second layer, under certain conditions the structural or the Schweigert-type instability arises. Each dust particle starts to act as an active agent and converts energy of the flowing plasmas into its own kinetic energy. The conversion of energy occurs only in the bilayer central region where the conditions for the development of instability are satisfied. Due to the conversion of energy, the central region melts. The peripheral subsystem where no instability is present is heated only by the heat flux from the central region. The flux of heat increases kinetic energy of particles at the periphery but it is not enough to melt the subsystem in a wide range of system parameters. In this case we observe solid-liquid phase co-existence that is supported by active agents located in the center of the structure.

\section*{\label{sec:theoryexp}Comparison between theory and experiment and discussion}

The employed simulation technique allows to obtain good agreement with the experimentally measured profile of kinetic temperature in the structure.  The profile is demonstrated in Fig. 8 for the estimated values of system parameters $Q=15000~e$, $\alpha=0.035$~cgs units, $\beta/\alpha=230$, $\gamma=1.8$~s$^{-1}$. This value of $\gamma$ equals to the experimental value calculated by the formula~(\ref{eqn:damping}) within the experimental error. The wake potential in this case is calculated for the value of ion flow velocity $v_{\rm fl} = v_{\rm B}$ which lies in the experimentally estimated range. Note that we do not expect the complete self-consistency of the developed multiscale approach due to the approximations employed in the numerical model. In order to improve the results, one should calculate the charge of each dust particle self-consistently with the dynamics of all dust particles. This is a resource-intensive task for the system of $2500$~dust grains.

Inspired by the observation of phase co-existence in our experiment, it is tempting to compare some of the features of our complex plasma system to those of the active matter models. In our case, the principal possibility of phase co-existence arises from the nonuniformity and thermodynamic openness of the system. Due to its finite size and action of the confinement, it has a bilayer structure in the central region and a monolayer structure at the periphery. As our MD simulations show, in the central region conditions for the development of wake-induced instabilities are satisfied. In presence of the instability, each charged dust particle in the central region can act as an active agent with the ion wake beneath such a dust particle serving as a source of input energy on a micro-scale. The interaction between the dust particle and the ion wake breaks the symmetry in the force balance, leading to a non-reciprocal interaction between the dust particles in the upper and lower layers. This non-reciprocal nature of interaction acts as a heat source, which helps to transfer the energy from the plasma species to the dust particles.

The energy gained by the dust particles in the form of kinetic energy is dissipated by the background neutrals. Akin to an active mater, the central part of the phase co-existing complex plasma system gets a continuous supply of heat energy from the particle beneath, which keeps this region melted. The peripheral region of the co-existing state, on the other hand, does not have such an energy source due to the absence of arising instability. As a result, it always remains in the crystalline phase. The heat energy also gets transferred from the center to the periphery along the longitudinal direction. The neutrals surrounding the crystalline state act as a heat sink and take away the excess energy from the particles in the crystalline phase.

Assuming that the action of the heat source in the single-layered peripheral section can be neglected, the heat transport equation in this region has the form ${\rm div}(\kappa \nabla T)=2\gamma n T k_B$, where $\kappa$ is the thermal conductivity, $n$ is the areal number density of dust particles. The experimentally measured temperature profile in the peripheral region is in a good agreement with the numerical solution of this equation. We estimate the value of the thermal diffusivity $\chi = \kappa/cn$, where $c$ is the specific heat of the dust subsystem. The estimated value is $\chi=5$~mm/s$^2$ which is close to the reported value $9$~mm/s$^2$ in a 2D dusty plasmas near melting \cite{nosenko2008heat}.

Our present experimental observations could serve as a paradigmatic example of a complex plasma system exhibiting dynamical features of an active matter with a stationary co-existence of solid and liquid phases. The proposed multiscale numerical model allows to simulate the process of phase co-existence in the system of active dust particles starting from the plasma microscale up to the macroscale where the dust system evolves.

\section*{\label{sec:Conclusion}Conclusion}

We experimentally observed a paradigmatic example of stable non-equilibrium coexistence of solid and fluid states in a complex plasmas system. The non-equilibrium nature of observed coexistence is confirmed by the measured non-uniform spatial profile of temperature in the structure. We conducted multiscale simulations of the system based on the kinetic equation for the calculation of electrostatic potential and on the Newton equation for the dynamics of dust particles. It is demonstrated that the coexistence is driven by the wake effect leading to arising of instabilities in the system. Due to structural non-uniformity under the action of the electrostatic trap, instabilities in the system develop locally. Our results prove the possibility of phase coexistence scenario in one-component plasmas which is often considered as a model system for classical condensed matter. We observe similarity between systems of active particles and dusty plasmas under certain conditions. The demonstrated coexistence of ordered and disordered states can be considered as the coexistence of a non-activated subsystem which is in thermal equilibrium with the neutral gas and a subsystem activated by per-particle energy exchange with surrounding plasma. We hope our experimental observations along with the numerical modelling results can trigger further investigations into this intriguing state of phase coexistence and the acquired `active' nature of a dusty plasma system under certain conditions.

\section*{Acknowledgements}

A.S. is thankful to the Indian National Science Academy (INSA) for their support under the INSA Senior Scientist Fellowship Scheme. The work of V. S. Nikolaev  was supported by the Foundation for the Advancement of Theoretical Physics and Mathematics “BASIS”. The simulations were conducted on Desmos and Fisher supercomputers at JIHT RAS.

\section*{Data availability statement}
The datasets used and/or analysed during the current study available from the corresponding author on reasonable request.

\section*{Author contributions statement}

M.G Hariprasad and P. Bandyopadhyay conceived the experimental set up. M.G Hariprasad, S. Arumugam, G. Arora and S. Singh conducted the experiment(s). M.G Hariprasad analysed the results. V. S. Nikolaev and D. A. Kolotinskii conducted analytical and numerical calculations. D. A. Kolotinskii calculated the potential distribution around dust particles. V. S. Nikolaev conducted and analyzed molecular dynamics simulations. A. Sen, P. Bandyopadhyay and A. V. Timofeev provided support, discussion and suggestions, and supervised the work.  All authors contributed to finalizing the manuscript.

%\bibliography{sample}

\begin{thebibliography}{10}
\urlstyle{rm}
\expandafter\ifx\csname url\endcsname\relax
  \def\url#1{\texttt{#1}}\fi
\expandafter\ifx\csname urlprefix\endcsname\relax\def\urlprefix{URL }\fi
\expandafter\ifx\csname doiprefix\endcsname\relax\def\doiprefix{DOI: }\fi
\providecommand{\bibinfo}[2]{#2}
\providecommand{\eprint}[2][]{\url{#2}}

\bibitem{wigner}
\bibinfo{author}{Wigner, E.}
\newblock \bibinfo{journal}{\bibinfo{title}{Effects of the electron interaction
  on the energy levels of electrons in metals}}.
\newblock {\emph{\JournalTitle{Trans. Faraday. Soc.}}}
  \textbf{\bibinfo{volume}{34}}, \bibinfo{pages}{678} (\bibinfo{year}{1938}).

\bibitem{activecolloids}
\bibinfo{author}{Ball, P.}
\newblock \bibinfo{journal}{\bibinfo{title}{Colloids get active}}.
\newblock {\emph{\JournalTitle{Nat. Mater.}}} \textbf{\bibinfo{volume}{12}},
  \bibinfo{pages}{696--696} (\bibinfo{year}{2013}).

\bibitem{dustyplasmas}
\bibinfo{author}{Fortov, V.}, \bibinfo{author}{Ivlev, A.},
  \bibinfo{author}{Khrapak, S.}, \bibinfo{author}{Khrapak, A.} \&
  \bibinfo{author}{Morfill, G.}
\newblock \bibinfo{journal}{\bibinfo{title}{Complex (dusty) plasmas: Current
  status, open issues, perspectives}}.
\newblock {\emph{\JournalTitle{Phys. Rep.}}} \textbf{\bibinfo{volume}{421}},
  \bibinfo{pages}{1--103} (\bibinfo{year}{2005}).

\bibitem{phasedusty}
\bibinfo{author}{Vaulina, O.}, \bibinfo{author}{Vladimirov, S.},
  \bibinfo{author}{Petrov, O.} \& \bibinfo{author}{Fortov, V.}
\newblock \bibinfo{journal}{\bibinfo{title}{Criteria of phase transitions in a
  complex plasma}}.
\newblock {\emph{\JournalTitle{Phys. Rev. Lett.}}}
  \textbf{\bibinfo{volume}{88}}, \bibinfo{pages}{245002}
  (\bibinfo{year}{2002}).

\bibitem{melzer2012phase}
\bibinfo{author}{Melzer, A.} \emph{et~al.}
\newblock \bibinfo{journal}{\bibinfo{title}{Phase transitions of finite dust
  clusters in dusty plasmas}}.
\newblock {\emph{\JournalTitle{Contrib. Plasma Phys.}}}
  \textbf{\bibinfo{volume}{52}}, \bibinfo{pages}{795--803}
  (\bibinfo{year}{2012}).

\bibitem{wavesdusty}
\bibinfo{author}{Piel, A.}, \bibinfo{author}{Klindworth, M.},
  \bibinfo{author}{Arp, O.}, \bibinfo{author}{Melzer, A.} \&
  \bibinfo{author}{Wolter, M.}
\newblock \bibinfo{journal}{\bibinfo{title}{Obliquely propagating dust-density
  plasma waves in the presence of an ion beam}}.
\newblock {\emph{\JournalTitle{Phys. Rev. Lett.}}}
  \textbf{\bibinfo{volume}{97}}, \bibinfo{pages}{205009}
  (\bibinfo{year}{2006}).

\bibitem{fortovmorfill}
\bibinfo{author}{Fortov, V.~E.} \& \bibinfo{author}{Morfill, G.}
\newblock \emph{\bibinfo{title}{Complex and Dusty Plasmas: from Laboratory to
  Space}} (\bibinfo{publisher}{CRC Press}, \bibinfo{year}{2009}).

\bibitem{polyakov2015synergetics}
\bibinfo{author}{Polyakov, D.}, \bibinfo{author}{Vasilyak, L.} \&
  \bibinfo{author}{Shumova, V.}
\newblock \bibinfo{journal}{\bibinfo{title}{Synergetics of dusty plasma and
  technological aspects of the application of cryogenic dusty plasma}}.
\newblock {\emph{\JournalTitle{Surface Engineering and Applied
  Electrochemistry}}} \textbf{\bibinfo{volume}{51}}, \bibinfo{pages}{143--151}
  (\bibinfo{year}{2015}).

\bibitem{transportdusty}
\bibinfo{author}{Fortov, V.} \emph{et~al.}
\newblock \bibinfo{journal}{\bibinfo{title}{Experimental study of the heat
  transport processes in dusty plasma fluid}}.
\newblock {\emph{\JournalTitle{Phys. Rev. E}}} \textbf{\bibinfo{volume}{75}},
  \bibinfo{pages}{026403} (\bibinfo{year}{2007}).

\bibitem{activematterreview}
\bibinfo{author}{Marchetti, M.~C.} \emph{et~al.}
\newblock \bibinfo{journal}{\bibinfo{title}{Hydrodynamics of soft active
  matter}}.
\newblock {\emph{\JournalTitle{Rev. Mod. Phys.}}}
  \textbf{\bibinfo{volume}{85}}, \bibinfo{pages}{1143} (\bibinfo{year}{2013}).

\bibitem{cellsintissues}
\bibinfo{author}{Alert, R.} \& \bibinfo{author}{Trepat, X.}
\newblock \bibinfo{journal}{\bibinfo{title}{Physical models of collective cell
  migration}}.
\newblock {\emph{\JournalTitle{Annu. Rev. Condens. Matter Phys.}}}
  \textbf{\bibinfo{volume}{11}}, \bibinfo{pages}{77--101}
  (\bibinfo{year}{2020}).

\bibitem{bacteriasuspensions}
\bibinfo{author}{Sokolov, A.} \& \bibinfo{author}{Aranson, I.~S.}
\newblock \bibinfo{journal}{\bibinfo{title}{Physical properties of collective
  motion in suspensions of bacteria}}.
\newblock {\emph{\JournalTitle{Phys. Rev. Lett.}}}
  \textbf{\bibinfo{volume}{109}}, \bibinfo{pages}{248109}
  (\bibinfo{year}{2012}).

\bibitem{flocksbirds}
\bibinfo{author}{Toner, J.} \& \bibinfo{author}{Tu, Y.}
\newblock \bibinfo{journal}{\bibinfo{title}{Flocks, herds, and schools: A
  quantitative theory of flocking}}.
\newblock {\emph{\JournalTitle{Phys. Rev. E}}} \textbf{\bibinfo{volume}{58}},
  \bibinfo{pages}{4828} (\bibinfo{year}{1998}).

\bibitem{ivlev2015statistical}
\bibinfo{author}{Ivlev, A.~V.} \emph{et~al.}
\newblock \bibinfo{journal}{\bibinfo{title}{Statistical mechanics where
  {New}ton's third law is broken}}.
\newblock {\emph{\JournalTitle{Phys. Rev. X}}} \textbf{\bibinfo{volume}{5}},
  \bibinfo{pages}{011035} (\bibinfo{year}{2015}).

\bibitem{januscolloids}
\bibinfo{author}{Vissers, T.}, \bibinfo{author}{Preisler, Z.},
  \bibinfo{author}{Smallenburg, F.}, \bibinfo{author}{Dijkstra, M.} \&
  \bibinfo{author}{Sciortino, F.}
\newblock \bibinfo{journal}{\bibinfo{title}{Predicting crystals of janus
  colloids}}.
\newblock {\emph{\JournalTitle{J. Chem. Phys.}}}
  \textbf{\bibinfo{volume}{138}}, \bibinfo{pages}{164505}
  (\bibinfo{year}{2013}).

\bibitem{khrapak2009basic}
\bibinfo{author}{Khrapak, S.} \& \bibinfo{author}{Morfill, G.}
\newblock \bibinfo{journal}{\bibinfo{title}{Basic processes in complex (dusty)
  plasmas: Charging, interactions, and ion drag force}}.
\newblock {\emph{\JournalTitle{Contrib. Plasma Phys.}}}
  \textbf{\bibinfo{volume}{49}}, \bibinfo{pages}{148--168}
  (\bibinfo{year}{2009}).

\bibitem{nosenko2010laser}
\bibinfo{author}{Nosenko, V.}, \bibinfo{author}{Ivlev, A.} \&
  \bibinfo{author}{Morfill, G.}
\newblock \bibinfo{journal}{\bibinfo{title}{Laser-induced rocket force on a
  microparticle in a complex (dusty) plasma}}.
\newblock {\emph{\JournalTitle{Phys. Plasmas}}} \textbf{\bibinfo{volume}{17}},
  \bibinfo{pages}{123705} (\bibinfo{year}{2010}).

\bibitem{janusdusty}
\bibinfo{author}{Nosenko, V.}, \bibinfo{author}{Luoni, F.},
  \bibinfo{author}{Kaouk, A.}, \bibinfo{author}{Rubin-Zuzic, M.} \&
  \bibinfo{author}{Thomas, H.}
\newblock \bibinfo{journal}{\bibinfo{title}{Active janus particles in a complex
  plasma}}.
\newblock {\emph{\JournalTitle{Phys. Rev. Res.}}} \textbf{\bibinfo{volume}{2}},
  \bibinfo{pages}{033226} (\bibinfo{year}{2020}).

\bibitem{petrovactive}
\bibinfo{author}{Arkar, K.}, \bibinfo{author}{Vasiliev, M.~M.},
  \bibinfo{author}{Petrov, O.~F.}, \bibinfo{author}{Kononov, E.~A.} \&
  \bibinfo{author}{Trukhachev, F.~M.}
\newblock \bibinfo{journal}{\bibinfo{title}{Dynamics of active brownian
  particles in plasma}}.
\newblock {\emph{\JournalTitle{Molecules}}} \textbf{\bibinfo{volume}{26}},
  \bibinfo{pages}{561} (\bibinfo{year}{2021}).

\bibitem{wakeeffect}
\bibinfo{author}{Lemons, D.}, \bibinfo{author}{Murillo, M.},
  \bibinfo{author}{Daughton, W.} \& \bibinfo{author}{Winske, D.}
\newblock \bibinfo{journal}{\bibinfo{title}{Two-dimensional wake potentials in
  sub-and supersonic dusty plasmas}}.
\newblock {\emph{\JournalTitle{Phys. Plasmas}}} \textbf{\bibinfo{volume}{7}},
  \bibinfo{pages}{2306--2313} (\bibinfo{year}{2000}).

\bibitem{wakeeffect1}
\bibinfo{author}{Vladimirov, S.~V.} \& \bibinfo{author}{Nambu, M.}
\newblock \bibinfo{journal}{\bibinfo{title}{Attraction of charged particulates
  in plasmas with finite flows}}.
\newblock {\emph{\JournalTitle{Phys. Rev. E}}} \textbf{\bibinfo{volume}{52}},
  \bibinfo{pages}{R2172} (\bibinfo{year}{1995}).

\bibitem{wakeeffect2}
\bibinfo{author}{Zhdanov, S.}, \bibinfo{author}{Ivlev, A.} \&
  \bibinfo{author}{Morfill, G.}
\newblock \bibinfo{journal}{\bibinfo{title}{Mode-coupling instability of
  two-dimensional plasma crystals}}.
\newblock {\emph{\JournalTitle{Phys. Plasmas}}} \textbf{\bibinfo{volume}{16}},
  \bibinfo{pages}{083706} (\bibinfo{year}{2009}).

\bibitem{matthews2020dust}
\bibinfo{author}{Matthews, L.~S.} \emph{et~al.}
\newblock \bibinfo{journal}{\bibinfo{title}{Dust charging in dynamic ion
  wakes}}.
\newblock {\emph{\JournalTitle{Phys. Plasmas}}} \textbf{\bibinfo{volume}{27}},
  \bibinfo{pages}{023703} (\bibinfo{year}{2020}).

\bibitem{piel2017molecular}
\bibinfo{author}{Piel, A.}
\newblock \bibinfo{journal}{\bibinfo{title}{Molecular dynamics simulation of
  ion flows around microparticles}}.
\newblock {\emph{\JournalTitle{Phys. Plasmas}}} \textbf{\bibinfo{volume}{24}},
  \bibinfo{pages}{033712} (\bibinfo{year}{2017}).

\bibitem{hutchinson2007computation}
\bibinfo{author}{Hutchinson, I.} \& \bibinfo{author}{Patacchini, L.}
\newblock \bibinfo{journal}{\bibinfo{title}{Computation of the effect of
  neutral collisions on ion current to a floating sphere in a stationary
  plasma}}.
\newblock {\emph{\JournalTitle{Phys. Plasmas}}} \textbf{\bibinfo{volume}{14}},
  \bibinfo{pages}{013505} (\bibinfo{year}{2007}).

\bibitem{miloch2010charging}
\bibinfo{author}{Miloch, W.}, \bibinfo{author}{Kroll, M.} \&
  \bibinfo{author}{Block, D.}
\newblock \bibinfo{journal}{\bibinfo{title}{Charging and dynamics of a dust
  grain in the wake of another grain in flowing plasmas}}.
\newblock {\emph{\JournalTitle{Phys. Plasmas}}} \textbf{\bibinfo{volume}{17}},
  \bibinfo{pages}{103703} (\bibinfo{year}{2010}).

\bibitem{melzer2014connecting}
\bibinfo{author}{Melzer, A.}
\newblock \bibinfo{journal}{\bibinfo{title}{Connecting the wakefield
  instabilities in dusty plasmas}}.
\newblock {\emph{\JournalTitle{Phys. Rev. E}}} \textbf{\bibinfo{volume}{90}},
  \bibinfo{pages}{053103} (\bibinfo{year}{2014}).

\bibitem{ivlev2017instabilities}
\bibinfo{author}{Ivlev, A.} \& \bibinfo{author}{Kompaneets, R.}
\newblock \bibinfo{journal}{\bibinfo{title}{Instabilities in bilayer complex
  plasmas: Wake-induced mode coupling}}.
\newblock {\emph{\JournalTitle{Phys. Rev. E}}} \textbf{\bibinfo{volume}{95}},
  \bibinfo{pages}{053202} (\bibinfo{year}{2017}).

\bibitem{torsion}
\bibinfo{author}{Nosenko, V.}, \bibinfo{author}{Zhdanov, S.~K.},
  \bibinfo{author}{Thomas, H.~M.}, \bibinfo{author}{Carmona-Reyes, J.} \&
  \bibinfo{author}{Hyde, T.~W.}
\newblock \bibinfo{journal}{\bibinfo{title}{Dynamics of spinning particle pairs
  in a single-layer complex plasma crystal}}.
\newblock {\emph{\JournalTitle{Phys. Rev. E}}} \textbf{\bibinfo{volume}{96}},
  \bibinfo{pages}{011201}, \doiprefix\url{10.1103/PhysRevE.96.011201}
  (\bibinfo{year}{2017}).

\bibitem{vortex_flows}
\bibinfo{author}{Uchida, G.}, \bibinfo{author}{Iizuka, S.},
  \bibinfo{author}{Kamimura, T.} \& \bibinfo{author}{Sato, N.}
\newblock \bibinfo{journal}{\bibinfo{title}{Generation of two-dimensional dust
  vortex flows in a direct current discharge plasma}}.
\newblock {\emph{\JournalTitle{Physics of Plasmas}}}
  \textbf{\bibinfo{volume}{16}}, \bibinfo{pages}{053707},
  \doiprefix\url{10.1063/1.3139252} (\bibinfo{year}{2009}).
\newblock \eprint{https://doi.org/10.1063/1.3139252}.

\bibitem{shock_coexistence}
\bibinfo{author}{Qiu, P.}, \bibinfo{author}{Sun, T.} \& \bibinfo{author}{Feng,
  Y.}
\newblock \bibinfo{journal}{\bibinfo{title}{Observation of the solid and liquid
  separation after the shock propagation in a two-dimensional yukawa solid}}.
\newblock {\emph{\JournalTitle{Physics of Plasmas}}}
  \textbf{\bibinfo{volume}{28}}, \bibinfo{pages}{113702},
  \doiprefix\url{10.1063/5.0067155} (\bibinfo{year}{2021}).
\newblock \eprint{https://doi.org/10.1063/5.0067155}.

\bibitem{melzer1996structure}
\bibinfo{author}{Melzer, A.} \emph{et~al.}
\newblock \bibinfo{journal}{\bibinfo{title}{Structure and stability of the
  plasma crystal}}.
\newblock {\emph{\JournalTitle{Phys. Rev. E}}} \textbf{\bibinfo{volume}{54}},
  \bibinfo{pages}{R46} (\bibinfo{year}{1996}).

\bibitem{schweigert1996alignment}
\bibinfo{author}{Schweigert, V.}, \bibinfo{author}{Schweigert, I.},
  \bibinfo{author}{Melzer, A.}, \bibinfo{author}{Homann, A.} \&
  \bibinfo{author}{Piel, A.}
\newblock \bibinfo{journal}{\bibinfo{title}{Alignment and instability of dust
  crystals in plasmas}}.
\newblock {\emph{\JournalTitle{Phys. Rev. E}}} \textbf{\bibinfo{volume}{54}},
  \bibinfo{pages}{4155} (\bibinfo{year}{1996}).

\bibitem{schweigert1998plasma}
\bibinfo{author}{Schweigert, V.}, \bibinfo{author}{Schweigert, I.},
  \bibinfo{author}{Melzer, A.}, \bibinfo{author}{Homann, A.} \&
  \bibinfo{author}{Piel, A.}
\newblock \bibinfo{journal}{\bibinfo{title}{Plasma crystal melting: A
  nonequilibrium phase transition}}.
\newblock {\emph{\JournalTitle{Phys. Rev. Lett.}}}
  \textbf{\bibinfo{volume}{80}}, \bibinfo{pages}{5345} (\bibinfo{year}{1998}).

\bibitem{couedel2009first}
\bibinfo{author}{Cou{\"e}del, L.} \emph{et~al.}
\newblock \bibinfo{journal}{\bibinfo{title}{First direct measurement of optical
  phonons in 2d plasma crystals}}.
\newblock {\emph{\JournalTitle{Phys. Rev. Lett.}}}
  \textbf{\bibinfo{volume}{103}}, \bibinfo{pages}{215001}
  (\bibinfo{year}{2009}).

\bibitem{couedel2011wave}
\bibinfo{author}{Cou{\"e}del, L.} \emph{et~al.}
\newblock \bibinfo{journal}{\bibinfo{title}{Wave mode coupling due to plasma
  wakes in two-dimensional plasma crystals: In-depth view}}.
\newblock {\emph{\JournalTitle{Phys. Plasmas}}} \textbf{\bibinfo{volume}{18}},
  \bibinfo{pages}{083707} (\bibinfo{year}{2011}).

\bibitem{prxnonuniformity}
\bibinfo{author}{Soni, V.}, \bibinfo{author}{G{\'o}mez, L.~R.} \&
  \bibinfo{author}{Irvine, W.~T.}
\newblock \bibinfo{journal}{\bibinfo{title}{Emergent geometry of inhomogeneous
  planar crystals}}.
\newblock {\emph{\JournalTitle{Phys. Rev. X}}} \textbf{\bibinfo{volume}{8}},
  \bibinfo{pages}{011039} (\bibinfo{year}{2018}).

\bibitem{hari_transition}
\bibinfo{author}{Hariprasad, M.~G.}, \bibinfo{author}{Bandyopadhyay, P.},
  \bibinfo{author}{Arora, G.} \& \bibinfo{author}{Sen, A.}
\newblock \bibinfo{journal}{\bibinfo{title}{Experimental observation of a
  first-order phase transition in a complex plasma monolayer crystal}}.
\newblock {\emph{\JournalTitle{Phys. Rev. E}}} \textbf{\bibinfo{volume}{101}},
  \bibinfo{pages}{043209} (\bibinfo{year}{2020}).

\bibitem{dpexcrystal}
\bibinfo{author}{Hariprasad, M.~G.}, \bibinfo{author}{Bandyopadhyay, P.},
  \bibinfo{author}{Arora, G.} \& \bibinfo{author}{Sen, A.}
\newblock \bibinfo{journal}{\bibinfo{title}{Experimental observation of a dusty
  plasma crystal in the cathode sheath of a dc glow discharge plasma}}.
\newblock {\emph{\JournalTitle{Physics of Plasmas}}}
  \textbf{\bibinfo{volume}{25}}, \bibinfo{pages}{123704}
  (\bibinfo{year}{2018}).

\bibitem{nikolaevtimofeev}
\bibinfo{author}{Nikolaev, V.} \& \bibinfo{author}{Timofeev, A.}
\newblock \bibinfo{journal}{\bibinfo{title}{Inhomogeneity of a harmonically
  confined {Yukawa} system}}.
\newblock {\emph{\JournalTitle{Physics of Plasmas}}}
  \textbf{\bibinfo{volume}{26}}, \bibinfo{pages}{073701}
  (\bibinfo{year}{2019}).

\bibitem{nikolaev2021nonhomogeneity}
\bibinfo{author}{Nikolaev, V.} \& \bibinfo{author}{Timofeev, A.}
\newblock \bibinfo{journal}{\bibinfo{title}{Nonhomogeneity of phase state in a
  dusty plasma monolayer with nonreciprocal particle interactions}}.
\newblock {\emph{\JournalTitle{Phys. Plasmas}}} \textbf{\bibinfo{volume}{28}},
  \bibinfo{pages}{033704} (\bibinfo{year}{2021}).

\bibitem{dpexii}
\bibinfo{author}{Arumugam, S.} \emph{et~al.}
\newblock \bibinfo{journal}{\bibinfo{title}{{DPEx-II}: a new dusty plasma
  device capable of producing large sized dc coulomb crystals}}.
\newblock {\emph{\JournalTitle{Plasma Sources Science and Technology}}}
  (\bibinfo{year}{2021}).

\bibitem{ikezi1986}
\bibinfo{author}{Ikezi, H.}
\newblock \bibinfo{journal}{\bibinfo{title}{Coulomb solid of small particles in
  plasmas}}.
\newblock {\emph{\JournalTitle{The Physics of Fluids}}}
  \textbf{\bibinfo{volume}{29}}, \bibinfo{pages}{1764--1766}
  (\bibinfo{year}{1986}).

\bibitem{vaulina_melting}
\bibinfo{author}{Vaulina, O.} \& \bibinfo{author}{Khrapak, S.}
\newblock \bibinfo{journal}{\bibinfo{title}{Scaling law for the fluid-solid
  phase transition in yukawa systems (dusty plasmas)}}.
\newblock {\emph{\JournalTitle{J. Exp. Theor. Phys}}}
  \textbf{\bibinfo{volume}{90}}, \bibinfo{pages}{287–289},
  \doiprefix\url{10.1134/1.559102} (\bibinfo{year}{2000}).

\bibitem{langevindynamics}
\bibinfo{author}{Knapek, C.~A.}, \bibinfo{author}{Ivlev, A.~V.},
  \bibinfo{author}{Klumov, B.~A.}, \bibinfo{author}{Morfill, G.~E.} \&
  \bibinfo{author}{Samsonov, D.}
\newblock \bibinfo{journal}{\bibinfo{title}{Kinetic characterization of
  strongly coupled systems}}.
\newblock {\emph{\JournalTitle{Phys. Rev. Lett.}}}
  \textbf{\bibinfo{volume}{98}}, \bibinfo{pages}{015001}
  (\bibinfo{year}{2007}).

\bibitem{voronoi}
\bibinfo{author}{Aurenhammer, F.}
\newblock \bibinfo{journal}{\bibinfo{title}{Voronoi diagrams -- a survey of a
  fundamental geometric data structure}}.
\newblock {\emph{\JournalTitle{ACM Comput. Surv.}}}
  \textbf{\bibinfo{volume}{23}}, \bibinfo{pages}{345--405}
  (\bibinfo{year}{1991}).

\bibitem{equilibrium}
\bibinfo{author}{Speck, T.}
\newblock \bibinfo{journal}{\bibinfo{title}{Coexistence of active {Brownian}
  disks: van der {Waals} theory and analytical results}}.
\newblock {\emph{\JournalTitle{Phys. Rev. E}}} \textbf{\bibinfo{volume}{103}},
  \bibinfo{pages}{012607} (\bibinfo{year}{2021}).

\bibitem{activematter}
\bibinfo{author}{Mandal, S.}, \bibinfo{author}{Liebchen, B.} \&
  \bibinfo{author}{L\"owen, H.}
\newblock \bibinfo{journal}{\bibinfo{title}{Motility-induced temperature
  difference in coexisting phases}}.
\newblock {\emph{\JournalTitle{Phys. Rev. Lett.}}}
  \textbf{\bibinfo{volume}{123}}, \bibinfo{pages}{228001}
  (\bibinfo{year}{2019}).

\bibitem{nosenko2008heat}
\bibinfo{author}{Nosenko, V.} \emph{et~al.}
\newblock \bibinfo{journal}{\bibinfo{title}{Heat transport in a two-dimensional
  complex (dusty) plasma at melting conditions}}.
\newblock {\emph{\JournalTitle{Phys. Rev. Lett.}}}
  \textbf{\bibinfo{volume}{100}}, \bibinfo{pages}{025003}
  (\bibinfo{year}{2008}).

\bibitem{instability1}
\bibinfo{author}{Schweigert, V.~A.}, \bibinfo{author}{Schweigert, I.~V.},
  \bibinfo{author}{Melzer, A.}, \bibinfo{author}{Homann, A.} \&
  \bibinfo{author}{Piel, A.}
\newblock \bibinfo{journal}{\bibinfo{title}{Plasma crystal melting: A
  nonequilibrium phase transition}}.
\newblock {\emph{\JournalTitle{Phys. Rev. Lett.}}}
  \textbf{\bibinfo{volume}{80}}, \bibinfo{pages}{5345--5348}
  (\bibinfo{year}{1998}).

\bibitem{instability2}
\bibinfo{author}{Schweigert, V.~A.}, \bibinfo{author}{Schweigert, I.~V.},
  \bibinfo{author}{Melzer, A.}, \bibinfo{author}{Homann, A.} \&
  \bibinfo{author}{Piel, A.}
\newblock \bibinfo{journal}{\bibinfo{title}{Alignment and instability of dust
  crystals in plasmas}}.
\newblock {\emph{\JournalTitle{Phys. Rev. E}}} \textbf{\bibinfo{volume}{54}},
  \bibinfo{pages}{4155--4166} (\bibinfo{year}{1996}).

\bibitem{Prx}
\bibinfo{author}{Ivlev, A.~V.} \emph{et~al.}
\newblock \bibinfo{journal}{\bibinfo{title}{Statistical mechanics where
  {Newton’s} third law is broken}}.
\newblock {\emph{\JournalTitle{Phys. Rev. X}}} \textbf{\bibinfo{volume}{5}},
  \bibinfo{pages}{011035} (\bibinfo{year}{2015}).

\bibitem{hutchinson2002ion}
\bibinfo{author}{Hutchinson, I.~H.}
\newblock \bibinfo{journal}{\bibinfo{title}{Ion collection by a sphere in a
  flowing plasma: I. {Q}uasineutral}}.
\newblock {\emph{\JournalTitle{Plasma Phys. Control. Fusion}}}
  \textbf{\bibinfo{volume}{44}}, \bibinfo{pages}{1953} (\bibinfo{year}{2002}).

\bibitem{hutchinson2006collisionless}
\bibinfo{author}{Hutchinson, I.}
\newblock \bibinfo{journal}{\bibinfo{title}{Collisionless ion drag force on a
  spherical grain}}.
\newblock {\emph{\JournalTitle{Plasma Phys. Control. Fusion}}}
  \textbf{\bibinfo{volume}{48}}, \bibinfo{pages}{185} (\bibinfo{year}{2006}).

\bibitem{miloch2012dust}
\bibinfo{author}{Miloch, W.} \& \bibinfo{author}{Block, D.}
\newblock \bibinfo{journal}{\bibinfo{title}{Dust grain charging in a wake of
  other grains}}.
\newblock {\emph{\JournalTitle{Phys. Plasmas}}} \textbf{\bibinfo{volume}{19}},
  \bibinfo{pages}{123703} (\bibinfo{year}{2012}).

\bibitem{changmai2020particle}
\bibinfo{author}{Changmai, S.} \& \bibinfo{author}{Bora, M.~P.}
\newblock \bibinfo{journal}{\bibinfo{title}{Particle-in-cell simulation of the
  effect of dust charge fluctuation on ion acoustic waves in a dusty plasma}}.
\newblock {\emph{\JournalTitle{Sci. Rep.}}} \textbf{\bibinfo{volume}{10}},
  \bibinfo{pages}{1--13} (\bibinfo{year}{2020}).

\bibitem{turq1977brownian}
\bibinfo{author}{Turq, P.}, \bibinfo{author}{Lantelme, F.} \&
  \bibinfo{author}{Friedman, H.~L.}
\newblock \bibinfo{journal}{\bibinfo{title}{Brownian dynamics: Its application
  to ionic solutions}}.
\newblock {\emph{\JournalTitle{J. Chem. Phys.}}} \textbf{\bibinfo{volume}{66}},
  \bibinfo{pages}{3039--3044} (\bibinfo{year}{1977}).

\bibitem{yukawaballs}
\bibinfo{author}{Henning, C.} \emph{et~al.}
\newblock \bibinfo{journal}{\bibinfo{title}{Ground state of a confined {Yukawa}
  plasma}}.
\newblock {\emph{\JournalTitle{Phys. Rev. E}}} \textbf{\bibinfo{volume}{74}},
  \bibinfo{pages}{056403} (\bibinfo{year}{2006}).

\bibitem{vaulina2009dynamics}
\bibinfo{author}{Vaulina, O.}, \bibinfo{author}{Koss, X.} \&
  \bibinfo{author}{Vladimirov, S.}
\newblock \bibinfo{journal}{\bibinfo{title}{The dynamics of formation of
  monolayer dust structures in a confining electric field}}.
\newblock {\emph{\JournalTitle{Phys. Scr.}}} \textbf{\bibinfo{volume}{79}},
  \bibinfo{pages}{035501} (\bibinfo{year}{2009}).

\bibitem{baumgartner2007shell}
\bibinfo{author}{Baumgartner, H.} \emph{et~al.}
\newblock \bibinfo{journal}{\bibinfo{title}{Shell structure of yukawa balls}}.
\newblock {\emph{\JournalTitle{Contrib. Plasma Phys.}}}
  \textbf{\bibinfo{volume}{47}}, \bibinfo{pages}{281--290}
  (\bibinfo{year}{2007}).

\bibitem{boning2008melting}
\bibinfo{author}{B{\"o}ning, J.} \emph{et~al.}
\newblock \bibinfo{journal}{\bibinfo{title}{Melting of trapped few-particle
  systems}}.
\newblock {\emph{\JournalTitle{Phys. Rev. Lett.}}}
  \textbf{\bibinfo{volume}{100}}, \bibinfo{pages}{113401}
  (\bibinfo{year}{2008}).

\bibitem{zhou2002distance}
\bibinfo{author}{Zhou, Y.}, \bibinfo{author}{Karplus, M.},
  \bibinfo{author}{Ball, K.~D.} \& \bibinfo{author}{Berry, R.~S.}
\newblock \bibinfo{journal}{\bibinfo{title}{The distance fluctuation criterion
  for melting: Comparison of square-well and {Morse} potential models for
  clusters and homopolymers}}.
\newblock {\emph{\JournalTitle{J. Chem. Phys.}}}
  \textbf{\bibinfo{volume}{116}}, \bibinfo{pages}{2323--2329}
  (\bibinfo{year}{2002}).

\bibitem{klumov2009structural}
\bibinfo{author}{Klumov, B.~A.} \& \bibinfo{author}{Morfill, G.}
\newblock \bibinfo{journal}{\bibinfo{title}{Structural properties of complex
  (dusty) plasma upon crystallization and melting}}.
\newblock {\emph{\JournalTitle{JETP Lett.}}} \textbf{\bibinfo{volume}{90}},
  \bibinfo{pages}{444--448} (\bibinfo{year}{2009}).

\end{thebibliography}

\end{document}